\documentclass[prl,aps,twocolumn,floatfix,longbibliography]{revtex4-2}
\usepackage{hyperref}
\usepackage[utf8]{inputenc}
\usepackage{natbib}
\usepackage{graphicx}
\usepackage{color}
\usepackage[tbtags]{amsmath}
\usepackage{amssymb}
\usepackage{gensymb}
\usepackage{float}

\begin{document}

\title{Andreev reflection at the altermagnet-superconductor interface}

\author{Micha{\l} Papaj}
\affiliation{Department of Physics, University of California, Berkeley, CA 94720, USA}

\begin{abstract}
Altermagnets are a new class of magnetic materials, which exhibit large spin splitting, but due to the combined spin and real space group symmetry protection maintain zero net macroscopic magnetization. Such a characteristic may prove them to be superior in applications in superconducting heterostructures and thus here we investigate the Andreev reflection at the altermagnet/superconductor interface. We compare and contrast altermagnets to other magnetic materials, revealing qualitative differences in the behavior of altermagnetic junction depending on the Fermi surface orientation. We study the resonant states arising in setups with strong tunneling barriers and show that sensitivity to non-magnetic disorder is also dependent on the orientation. Our results provide a building block for altermagnetic superconducting heterostructures such as Josephson $\pi$ junctions with superior properties.
\end{abstract}

\maketitle

\textit{Introduction}.---
The interplay between magnetism and superconductivity is the source of some of the most fascinating phenomena in condensed matter physics \cite{buzdinProximityEffectsSuperconductorferromagnet2005, bergeretOddTripletSuperconductivity2005, blamireInterfaceSuperconductivityMagnetism2014, linderSuperconductingSpintronics2015, pawlakMajoranaFermionsMagnetic2019, flensbergEngineeredPlatformsTopological2021}. In particular, the exchange field and associated spin splitting can lead to the emergence of spatially oscillating superconducting order parameters of Fulde-Ferrell-Larkin-Ovchinnikov states \cite{fuldeSuperconductivityStrongSpinExchange1964, larkinNonuniformStateSuperconductors1964} or topological superconductivity with effective $p+ip$ pairing and the presence of Majorana zero modes \cite{sauGenericNewPlatform2010, chungTopologicalSuperconductingPhase2011, duckheimAndreevReflectionNoncentrosymmetric2011, nakosaiTwodimensionalWaveSuperconducting2013, papajCreatingMajoranaModes2021}. Therefore, understanding the behavior of the interface between the magnetic materials and superconductors is of utmost theoretical and experimental interest.

At the microscopic level, the fundamental role in the physics of normal-superconductor interface is played by the competition of the Andreev and normal reflection processes \cite{andreevThermalConductivityIntermediate1964, klapwijkProximityEffectAndreev2004, dagheroProbingMultibandSuperconductivity2010, leeRecentProgressProbing2016} . When a spin up electron traveling through a material in normal state encounters a superconducting interface, it can either undergo normal reflection as spin up electron, or Andreev reflection as a spin down hole, with analogous process possible for the opposite spins as well. In contrast to normal reflection, during the Andreev reflection, a Cooper pair with charge $2e$ is transferred into the superconducting condensate, leading to the enhancement of conductance compared to the normal state. The effect of such processes on the transport properties of the junction can be understood through the elegant formalism of Blonder, Tinkham, and Klapwijk (BTK) \cite{blonderTransitionMetallicTunneling1982, riedelCurrentvoltageRelationNormalmetal1993, chaudhuriAndreevResonancesCurrentvoltage1995}.  When time-reversal symmetry is preserved and no spin-splitting occurs, quasiparticles of both spin orientations are degenerate and thus their reflection processes are equivalent. However, when a magnetic material with broken time-reversal symmetry is introduced, this equivalence is broken, leading to the suppression of Andreev reflection for one spin direction, but not the other. This asymmetry lies at the core of the novel properties of the magnetic junctions \cite{buzdinProximityEffectsSuperconductorferromagnet2005, bergeretOddTripletSuperconductivity2005, blamireInterfaceSuperconductivityMagnetism2014}.

So far, most of the magnetic-superconductor interfaces used either ferromagnets \cite{dejongAndreevReflectionFerromagnetSuperconductor1995,  demlerSuperconductingProximityEffects1997, falkoAndreevReflectionsMagnetoresistance1999, kashiwayaSpinCurrentFerromagnetinsulatorsuperconductor1999, mazinProbingSpinPolarization2001, hiraiTemperatureDependenceSpinpolarized2003, asanoOddfrequencyPairsJosephson2007, asanoJosephsonEffectDue2007, yokoyamaManifestationOddfrequencySpintriplet2007, robinsonControlledInjectionSpinTriplet2010, annunziataChargeSpinTransport2011, nessSupercurrentDecayBallistic2022}, characterized by large, macroscopic magnetization, or antiferromagnets \cite{bobkovaSpinDependentQuasiparticleReflection2005, andersenEnsuremathEnsuremathPi2006, komissinskiyJosephsonEffectHybrid2007, johnsenMagneticControlSuperconducting2021, bobkovEelProximityEffect2022, jakobsenElectricalThermalTransport2020} with magnetic components compensated on a microscopic level. However, recently a new class of magnetic materials with large spin-splitting, but with zero net magnetization protected by a combination of spin and real space symmetries called altermagnets was discovered and characterized \cite{smejkalEmergingResearchLandscape2022, smejkalConventionalFerromagnetismAntiferromagnetism2022, smejkalGiantTunnelingMagnetoresistance2022, mazinPredictionUnconventionalMagnetism2021, reichlovaMacroscopicTimeReversal2020, yangSymmetryInvariantsMagnetically2023}. As altermagnets are expected to be abundant both in two and three dimensional crystals, it is intriguing to explore new features that could be unique to this new magnetic phase. For example, the lack of net magnetization may prove to be of a great benefit in the superconducting heterostructures as the stray fields arising from the ferromagnetic materials are largely detrimental to superconducting pairing and often require external compensation and carefully designed experiments \cite{blamireInterfaceSuperconductivityMagnetism2014}. One particular example of structure that would benefit from this are Josephson $\pi$ junctions \cite{buzdinCriticalcurrentOscillationsFunction1982, ryazanovCouplingTwoSuperconductors2001} and very recently, presence of such a regime was proposed in junctions with altermagnetic barriers \cite{ouassouJosephsonEffectAltermagnets2023, zhangFinitemomentumCooperPairing2023, beenakkerPhaseshiftedAndreevLevels2023}. Moreover, very recently proposals for topological superconductivity in proximitized altermagnets have been made \cite{ghorashiAltermagneticRoutesMajorana2023, zhuTopologicalSuperconductivityTwoDimensional2023}. These and other possible advantages of altermagnets are encouraging to explore their superconducting junctions in more depth.

In this work we investigate the transport properties of the altermagnet/superconductor interface using the BTK formalism. Such interfaces are the basic building blocks for understanding the interplay between superconductivity and magnetism. We show that depending on the orientation of the spin-split altermagnet Fermi surface with regards to the interface plane, transport properties can either resemble qualitatively that of a ferromagnet with Andreev reflection suppressed or retain the enhanced in-gap conductance characteristic of a non-magnetic junction. This is linked to the different availability of electron and hole scattering states of opposite spins for different Fermi surface orientations. We then investigate the impact of a potential barrier at the interface and within the altermagnet, revealing that depending on the altermagnet orientation a host of resonant states that enhance conductance can be present, with their energy dependent on the barrier-interface distance. Finally, we also investigate how these properties behave in the presence of non-magnetic disorder, showing that altermagnet orientation also affects the sensitivity of the junction transport properties to random impurities. These findings highlight the attractiveness of applying altermagnets in novel magnetic superconducting heterostructures.

\textit{Model}.---
To characterize the normal and Andreev reflection processes at the altermagnet/superconductor (AM/SC) interface we use the Bogoliubov-de Gennes Hamiltonian to describe the electron and hole excitations of the system:

\begin{equation}
\label{eq:Hamiltonian}
    \begin{pmatrix}
        H(\mathbf{r}) & \Delta(\mathbf{r}) \\
        \Delta^\dagger(\mathbf{r}) & - H^*(\mathbf{r}) \\
    \end{pmatrix} \Psi(\mathbf{r}) = E \Psi(\mathbf{r})
\end{equation}
where $\Psi(\mathbf{r}) = (u_\uparrow, u_\downarrow, v_\uparrow, v_\downarrow)^T$ is the Nambu spinor with $u_\sigma$ and $v_\sigma$ electron and hole components of spin $\sigma$, respectively.

We consider a 2D system with the interface separating the altermagnet and the superconductor at $x=0$. In such a case, the components of the Hamiltonian are \cite{smejkalEmergingResearchLandscape2022}:
\begin{align}
    H(\mathbf{r}) = H_{AM}\,\theta(-&x) + H_{SC}\,\theta(x) \notag \\
    H_{AM} = t_0(k_x^2 + k_y^2) + (t_{J1}(k_y^2&-k_x^2) + 2 t_{J2} k_x k_y)\sigma_z - \mu \notag\\
    H_{SC} = t_{SC}(k_x^2 &+ k_y^2) - \mu_{SC}
\end{align}
with momenta $k_{x,y} = -i\partial_{x,y}$ understood as the differential operators in a given direction. We allow for the change of both the effective mass and the chemical potential between the altermagnet and the superconductor in the formulas, with such a mismatch leading to the decrease in the transparency of the interface, effectively forming a tunneling barrier. Introducing such a step change at the interface requires additional care when considering the exact form of the Hamiltonian expressed in terms of differential operators \cite{vonroosPositiondependentEffectiveMasses1983, morrowModelEffectivemassHamiltonians1984}, which is explored in more detail in the Supplemental Material \cite{SM}. Including both $t_{J1}$ and $t_{J2}$ terms in the altermagnetic phase enables us to consider the two symmetric orientations of the spin-split Fermi surface and investigate its impact on possible reflection processes. Such a formulation also enables considering arbitrary orientations of the Fermi surface, with the rotation angle determined through $t_{J1} \cos(2\theta_k) + t_{J2} \sin(2\theta_k) = \sqrt{t_{J1}^2 + t_{J2}^2} \cos(2\theta_k + \arctan(-t_{J2}/t_{J1}))$ by the relative ratio $t_{J2}/t_{J1}$, with $\theta_k$ being the polar angle in momentum space. For the superconductor, we assume standard $s$-wave pairing with $\Delta(\mathbf{r}) = \Delta i\sigma_y \theta(x)$. For the purpose of comparison with ferromagnetic junction, to introduce spin splitting we use a exchange energy term in the normal state Hamiltonian $H_Z = B_Z \theta(-x) \sigma_z$. We also consider the presence of a tunneling barrier within the altermagnet or at the interface in a form of delta function potential $H_B = U_B\delta(x+L)$, where $L>0$ is the barrier position and $U_B$ is its strength parameter.

We also consider the lattice counterpart of model \eqref{eq:Hamiltonian} for the purpose of numerical calculations, enabling both direct comparison between the analytical and simulation results, as well as studying the impact of random disorder. The model is discretized on a square lattice with nearest and next nearest neighbor hoppings. In calculations with disorder, we model it as random non-magnetic potential chosen independently on each lattice site from a uniform distribution in the range $[-U_0 / 2, U_0 / 2]$, where we call $U_0$ the disorder strength. In a geometry of a finite width ribbon, two leads (altermagnetic and superconducting) are attached at opposite ends and scattering matrix of the system is computed using Kwant code \cite{GrothKwantsoftwarepackage2014}. From the scattering matrix, conductance of the interface is obtained by summing over the BTK conductance of each scattering mode.

\begin{figure}
    \centering
    \includegraphics[width=0.99\columnwidth]{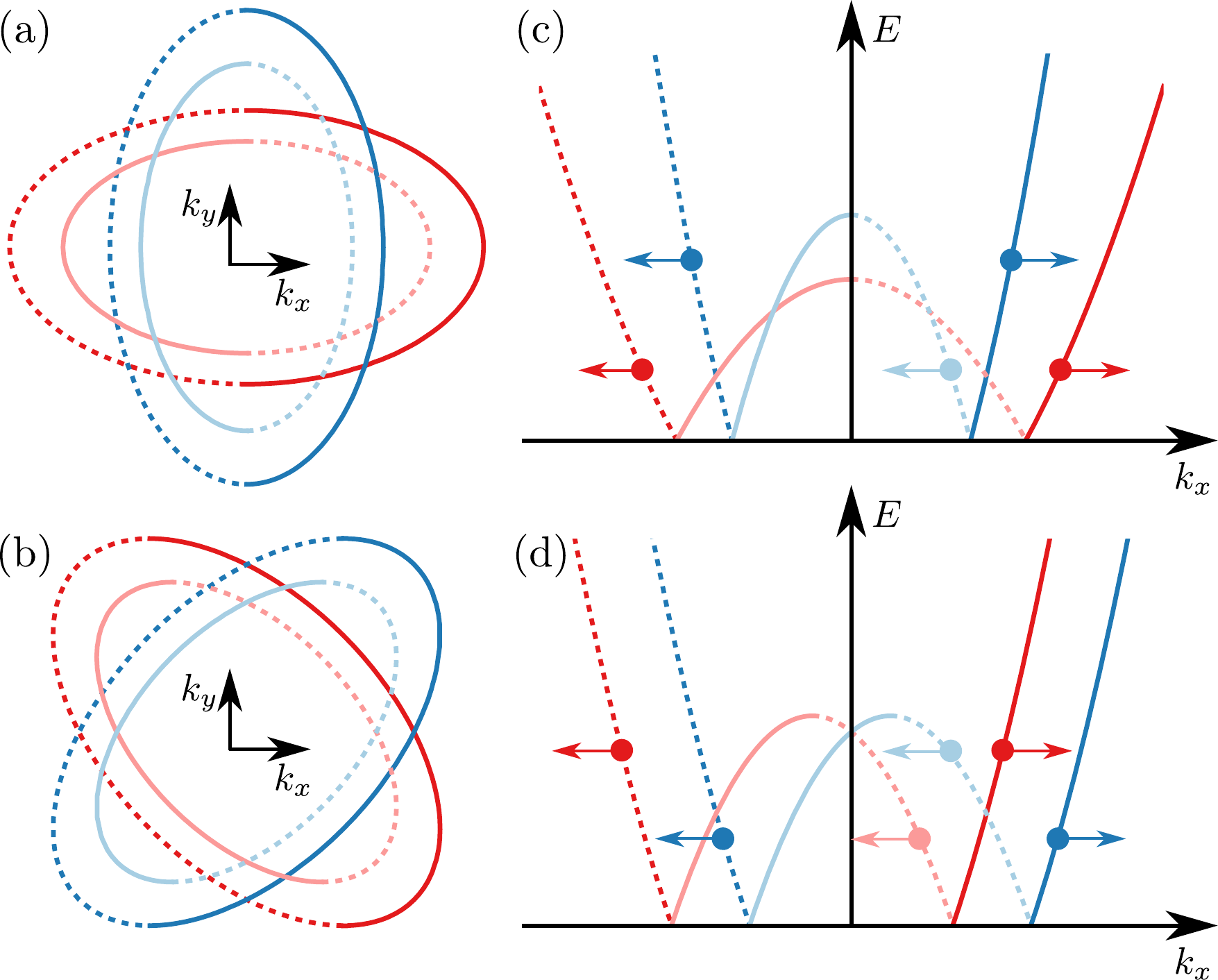}
    \caption{Quasiparticle excitations of Hamiltonian \eqref{eq:Hamiltonian} for (a,c) $t_{J1}>0, t_{J2}=0$ and (b,d) $t_{J1}=0, t_{J2}>0$.  (a) The constant energy contours for $E>0$ are ellipses with the major axes either parallel or perpendicular to the AM/SC interface. (b) The constant energy contours for $E>0$ are ellipses with the major axes at $\pi/4$ angle to the AM/SC interface. (c) Quasiparticle dispersion $E(k_x)$ for $k_y>0$ contains an energy range in which only normal reflection is possible for spin-down electrons. (d) Andreev and normal reflection are possible in the same energy range for both apin up and down electrons. In all panels red and blue correspond to spin up and down, dark and light to electron and hole, and solid and dashed to group velocity $v_x >0$ and $v_x<0$, respectively.}
    \label{fig:FS_dispersion}
\end{figure}

\textit{Excitations and reflection processes in altermagnet}.---
With such a Hamiltonian describing the interface we can first examine the electron and hole excitations within the altermagnet for the two orientations of the spin-splitting direction. We will consider two separate cases, with either $t_{J1}$ or $t_{J2}$ being non-zero. The constant energy contours at $E>0$ for electron and hole excitations in both of these scenarios are presented in Fig.~\ref{fig:FS_dispersion}(a,b). While in both cases the contours are elliptical, the orientation of these ellipses for spin up and down quasiparticles with respect to the AM/SC interface is different. For $t_{J1}>0$ the ellipse major axes are perpendicular to the interface for spin up and parallel to the interface for spin down particles, while for $t_{J2}>0$ case both major axes are at $\pi/4$ angle to the interface. This translates to a different set of restrictions on the possibility of Andreev reflection, during which the electron gets reflected as a hole of opposite spin and vice versa. In the first case, there is a range of $k_y$ for which a state with a given spin does not have a hole counterpart with the opposite spin, necessitating a purely normal reflection when the energies under consideration are within superconducting gap, similarly to the behavior in ferromagnet/superconductor junction. In contrast, in the second case, the energy range for which electrons and holes of opposite spin can coexist is the same for spin up and down particles. This means that Andreev reflection is always possible and the precise split between Andreev and normal processes depends on the details of wave length and effective mass mismatch across the interface. This difference is clearly visible in excitation dispersion presented in Fig.~\ref{fig:FS_dispersion}(c, d) for a $k_y>0$ in both of the analyzed cases. In panel \ref{fig:FS_dispersion}(c), when $t_{J1}>0$, there is an energy region within which spin down electron can only be reflected as another spin down electron moving in the opposite direction. In the same energy window, a spin up electron can be reflected as spin down hole as well as the spin up electron. However, in panel \ref{fig:FS_dispersion}(d) the dispersion is symmetrical for both spin up and down particles, and no energy window within which only normal reflection is possible exists. This will have direct consequences for the conductance across the SC/AM interface as we discuss below. While the model considered here is 2D, the discussion above would straightforwardly generalize to a 3D case, where an additional degree of freedom would arise that could be used to orient the altermagnet Fermi surface with respect to the interface between the materials.

\textit{Solution of the scattering problem}.---
To obtain the transport properties of the AM/SC junction, we solve the scattering problem with the wave function matching condition at the interface:
\begin{equation}
\label{eq:matching}
\begin{split}
    \Psi(x=0^-) &= \Psi(x=0^+) \\
    \tau_z\left(t_0 - t_{J1} \sigma_z \right) \partial_x \Psi \big|_{0^-} + t_{J2} &\tau_z\sigma_z \partial_y\Psi\big|_{0^-} = t_{SC} \tau_z \partial_x \Psi\big|_{0^+}
\end{split}
\end{equation}
with $\tau_z$ being the Pauli matrix in particle-hole space. The second condition, as derived in the Supplemental Material \cite{SM} is equivalent to the conservation of the quasiparticle current across the interface. Since we neglect the spin-orbit coupling and thus spin remains a good quantum number within the altermagnet, the matching condition splits into two sets of equations for spin up and down particles. In each set we consider an incoming and normal reflected electron states, and Andreev reflected hole within the altermagnet. Inside of the superconductor the wave function consists of electron- and hole-like quasiparticles, which are exponentially decaying for $E<\Delta$ and propagating through the superconductor above the gap. Solution of these equations yields the amplitudes for normal ($r_{N\sigma}$) and Andreev reflection ($r_{A\sigma}$) processes for both spin up and spin down electrons:
\begin{equation}
    r_{A\sigma} = \frac{2 \sigma \tilde{k}_{F,SC} \sqrt{\tilde{k}_{e\sigma}\tilde{k}_{h\bar{\sigma}}}}{\tilde{k}_{F,SC}(\tilde{k}_{e\sigma}+\tilde{k}_{h\bar{\sigma}})\tilde{\epsilon}+(\tilde{k}^{2}_{F,SC}+\tilde{k}_{e\sigma}\tilde{k}_{h\bar{\sigma}})\sqrt{\tilde{\epsilon}^2-1}} \\
\end{equation}
\begin{equation}
    r_{N\sigma} = \frac{\tilde{k}_{F,SC}(\tilde{k}_{e\sigma}-\tilde{k}_{h\bar{\sigma}})\tilde{\epsilon}+(\tilde{k}_{e\sigma}\tilde{k}_{h\bar{\sigma}}-\tilde{k}^{2}_{F,SC})\sqrt{\tilde{\epsilon}^2-1}}{\tilde{k}_{F,SC}(\tilde{k}_{e\sigma}+\tilde{k}_{h\bar{\sigma}})\tilde{\epsilon}+(\tilde{k}^{2}_{F,SC}+\tilde{k}_{e\sigma}\tilde{k}_{h\bar{\sigma}})\sqrt{\tilde{\epsilon}^2-1}}
\end{equation}
where we have defined $\tilde{\epsilon} = E/\Delta$ and $\sigma=\pm1$ for spin up and down. We also have:
\begin{equation}
\begin{aligned}
    &\tilde{k}_{F,SC} = t_{SC} \sqrt{\mu_{SC} / t_{SC} - k_y^2} \\
    &\tilde{k}_{e\sigma} = (t_0\mp t_{J1})\sqrt{\frac{\mu + \Delta \tilde{\epsilon}}{t_0\mp t_{J1}} - k_y^2\frac{t_0^2-t_{J1}^2-t_{J2}^2}{(t_0\mp t_{J1})^2}} \\
    &\tilde{k}_{h\sigma} = (t_0\mp t_{J1})\sqrt{\frac{\mu - \Delta \tilde{\epsilon}}{t_0\mp t_{J1}} - k_y^2\frac{t_0^2-t_{J1}^2-t_{J2}^2}{(t_0\mp t_{J1})^2}}
\end{aligned}
\end{equation}
Here the upper sign corresponds to spin up and lower sign to spin down component. In deriving these formulas we have assumed that the wave vector of the quasiparticle within the superconductor is independent of energy. A similar set of reflection coefficients can also be derived for the scenario with the barrier present ($U_B \neq 0$) as shown in the Supplemental Material \cite{SM}. In both of these cases, since the wave functions are normalized to carry a unit quasiparticle current, for $E < \Delta$ we have $|r_{A\sigma}|^2 + |r_{N\sigma}|^2=1$ as required by unitarity of the scattering matrix.

\begin{figure}
    \centering
    \includegraphics[width=0.99\columnwidth]{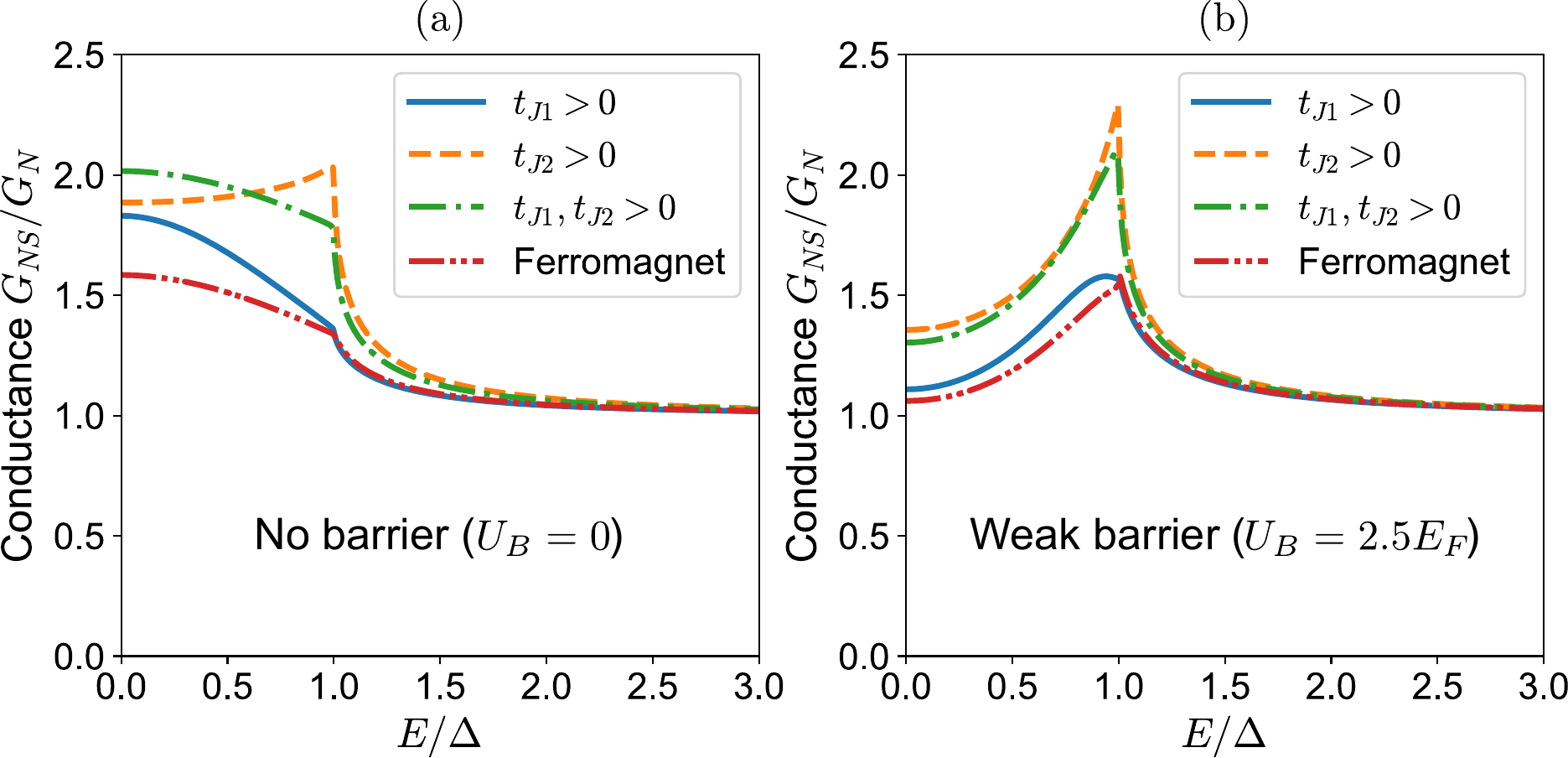}
    \caption{Normalized conductance of altermagnet/superconductor interface obtained from Eq.~\eqref{eq:BTK_conductance} for three different orientations of altermagnetic Fermi surface. In all cases, $t_0 =4$, $\mu=\mu_{SC}=0.25$, $\Delta = 0.001$. Blue lines are calculated for $t_{J1}=3$, orange lines for $t_{J2}=3$, green lines for $t_{J1}=t_{J2}=2.12$, and red lines for $B_Z=0.15$. (a) Perfect interface with no tunneling barrier. (b) Interface with a weak tunneling barrier.}
    \label{fig:interface_cond}
\end{figure}

\begin{figure*}
    \centering
    \includegraphics[width=0.999\linewidth]{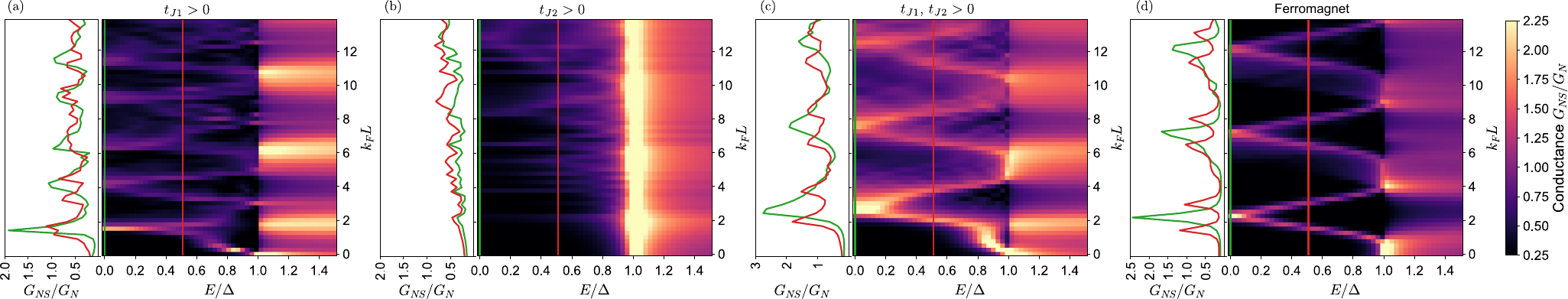}
    \caption{Conductance of interface with barrier placed distance $L$ away from the altermagnet/superconductor boundary at energy $E$, obtained using numerical simulation with system width $W=301$ lattice sites. In each case the vertical side panels present the line cut of the color maps at $E=0$ (green) and $E=\Delta/2$ (red). We scale the distance by $k_F = \sqrt{\mu/t_0}$. In all panels, $t_0 =4$, $\mu=\mu_{SC}=0.25$, $\Delta = 0.001$. (a) Altermagnet with $t_{J1}=3$, $t_{J2}=0$. (b) Altermagnet with $t_{J1}=0$, $t_{J2}=3$. (c) Altermagnet with $t_{J1}=t_{J2}=2.12$. (c) Ferromagnet with $B_Z = 0.15$. }
    \label{fig:barrier_distance}
\end{figure*}

With the reflection amplitudes derived, we can now employ the BTK formalism in temperature $T=0$ limit to obtain the conductance across the interface. In each case we normalize the conductance with respective conductance of a normal state interface, characterized by the normal reflection amplitude $r_{N0\sigma}$:
\begin{equation}
\label{eq:BTK_conductance}
    G_{NS}/G_N = \frac{\sum_\sigma \int dk_y \left(1 - |r_{N\sigma}|^2 + |r_{A\sigma}|^2 \right)}{\sum_\sigma \int dk_y \left(1 - |r_{N0\sigma}|^2\right)}
\end{equation}
The limits of integration are established by the extent of the constant energy contour ellipses for given $E$ and are:
\begin{equation}
    k_{y,e\sigma}^\mathrm{max} = \sqrt{\frac{(t_0 \mp t_{J1})(\mu + E)}{t_0^2-t_{J1}^2-t_{J2}^2}}, k_{y,h\sigma}^\mathrm{max} = \sqrt{\frac{(t_0 \mp t_{J1})(\mu - E)}{t_0^2-t_{J1}^2-t_{J2}^2}}
\end{equation}
The maximum extent of hole energy contour of a given spin $ k_{y,h\sigma}^\mathrm{max}$ limits the extent of Andreev reflection of the electron of opposite spin, beyond which the electron gets fully normally reflected for energies within the superconducting gap. When evaluating the integral we also include a small imaginary part $\eta$ in energy $E \rightarrow E + i\eta$ to provide regularization.

\textit{Interface conductance}.---
With the formalism for the interface conductance established, we can now investigate the impact of different orientations of altermagnet Fermi surface on the transport properties of the junction. We first consider a perfect interface with no barrier and the same chemical potential within the altermagnet and superconductor as presented in Fig.~\ref{fig:interface_cond}(a). Here the difference in the range of $k_y$ for which Andreev reflection is possible for the two orientations of the altermagnet is immediately obvious. While in the $t_{J1}$ case Andreev reflection becomes suppressed with decreasing conductance when energy increases towards the superconducting gap edge, in the $t_{J2}$ case the Andreev reflection leads to increased conductance with maximum at $E=\Delta$. When compared to the ferromagnetic interface, the $t_{J1}$ bears qualitative resemblance to it in the energy dependence of conductance, with nevertheless stronger Andreev reflection at $E=0$. In contrast, $t_{J2}$ altermagnet behaves more similarly to a non-magnetic system with imperfect Andreev reflection resulting from the barrier due to the Fermi wavevector mismatch. In the intermediate case, when $t_{J1}=t_{J2}>0$, the system mirrors the $t_{J1}$ energy dependence while also presenting increased conductance of the $t_{J2}$ case.

We then introduce a delta function barrier at the AM/SC interface and present the results for intermediate strength barrier in Fig.~\ref{fig:interface_cond}(b). There, the similarity between the $t_{J1}$ orientation and the ferromagnet is further strengthened, with the $E=0$ conductance advantage of the altermagnet diminished and the curves following a similar dependence. In contrast, the $t_{J2}$ and the mixed cases retain their conductance enhancement throughout the superconducting gap, and in particular at the gap edge. In all of the cases however, the behavior outside of the superconducting gap is largely equivalent. Finally, with a very strong barrier, all of the conductances collapse onto one dependence, which now results in probing the density of states of the superconductor, with resonant peaks at the gap edge and no conductance within the gap.

The delta function barrier can also be placed at some distance $L$ away from the interface within the altermagnet. This can lead to a formation of resonant states trapped between the barrier and interface through combination of normal and Andreev reflection processes. Conductance of such setup for varying $L$ and energy is presented in Fig.~\ref{fig:barrier_distance} for each of the previously analyzed altermagnetic cases and a ferromagnet for comparison, with the slices along the distance axis at $E=0$ and $E=0.5\Delta$ presented in the side panels. This again illustrates the qualitative differences in the behavior of AM/SC interface, dependent on the orientation of the AM Fermi surface. In panel Fig.~\ref{fig:barrier_distance}(a) we demonstrate the presence of the aforementioned resonant states within the gap in the $t_{J1}$ case, where they appear as interface conductance peaks at $E=0$ and $E=\Delta$ for barrier distance $L$ determined by the electron and hole wave vectors within the altermagnet. These peaks then split into pairs when the barrier distance is changed, appearing at intermediate energies $0 < E < \Delta$. Strong oscillations of conductance occur also for energies outside of the gap, but the position of the enhancement maximum no longer disperses with barrier distance. While similar resonant states are present in the ferromagnetic case (Fig.~\ref{fig:barrier_distance}(d)), the oscillations for energies above the gap are not nearly as pronounced. In contrast to this behavior, in the $t_{J2}$ case the resonant states are absent and the conductance within the gap slowly increases for larger barrier distances, with some noise imposed over this trend due to the higher order interference of multiple paths between the barrier and the interface. Moreover, in this case the conductance above the gap is uniform across all the barrier distances, with a sharp peak at the gap edge more akin to the standard normal-superconductor interface with a barrier. Finally, the mixed case with $t_{J1}=t_{J2}$ is again more similar to the $t_{J1}$ case, with the conductance behavior presenting the resonant states within the gap and strong conductance oscillations above the gap, albeit weaker than in a purely $t_{J1}$ system.

\begin{figure}
    \centering
    \includegraphics[width=0.99\columnwidth]{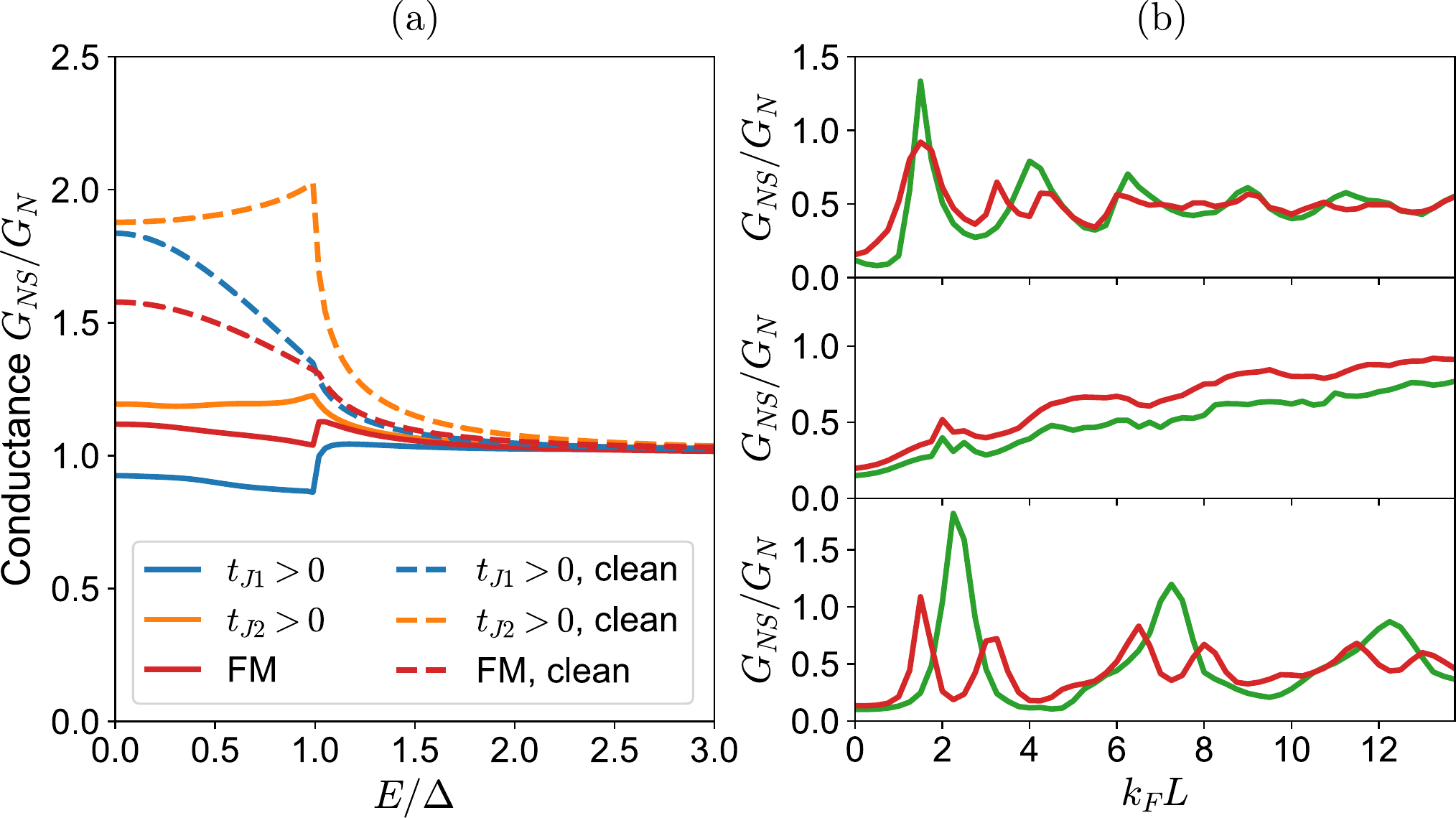}
    \caption{Conductance of altermagnet/superconductor interface obtained from numerical simulation with disorder included. The solid lines in all panels are conductance results averaged over 50 disorder realizations. (a) Disordered junction ($U_0 = 1$) with no delta function barrier in the system of width $W=201$ lattice sites and other parameters as in Fig.~\ref{fig:interface_cond}(a). Dashed lines show results for clean system. (b) Disordered junction with delta barrier placed distance $L$ away from the interface, with disorder between the barrier and the interface. The results correspond to line cuts presented in side panels of Fig.~\ref{fig:barrier_distance}(a,b,d), with green line at $E=0$, red line at $E=0.5\Delta$, and the same model parameters.}
    \label{fig:disorder}
\end{figure}

\textit{Impact of disorder}.---
Using the lattice model allows us to also investigate the impact of disorder on the interface conductance. We first consider the scenario where the altermagnet region contains random onsite disorder, but otherwise does not include a delta function barrier, with the results presented in Fig.~\ref{fig:disorder}(a). Both orientations of the altermagnet and the ferromagnet show similar response to disorder, but the decrease of conductance throughout the gap is the largest in $t_{J1}$ case. The behavior of the magnetic materials in this case is different than the one of normal spin-degenerate metal, as in that situation the effect of disorder can be compared to the effect of a tunneling barrier, with a strong resonant peak at the gap edge and quick decay of conductance within the gap. Here, however, the conductance remains stable throughout the gap. Comparatively, for the same strength of disorder, the $t_{J2}$ orientation of the altermagnet suffers less from the presence of disorder, with the conductance flat within the gap as well. In panel \ref{fig:disorder}(b) we then include a variable position delta barrier and show the dependence of interface conductance on the barrier distance at $E=0$ and $E=0.5\Delta$, the same as the side panels of Fig.~\ref{fig:barrier_distance}, with disorder included between the barrier and the interface. While in the $t_{J2}$ case the disorder does not change the results qualitatively, both $t_{J1}$ and the ferromagnet exhibit suppression of the conductance oscillation amplitude as the barrier distance increases.

\textit{Summary and outlook}.---
In conclusion, we have investigated the conductance of AM/SC junction, characterizing its dependence on the altermagnet Fermi surface orientation with respect to the interface. We have shown that while the $t_{J1}$ case is qualitatively similar to a ferromagnet, the $t_{J2}$ altermagnet does not suffer from suppressed Andreev reflection and thus preserves enhanced conductance within the superconducting gap. Our results highlight the importance of determining the crystalline orientation of altermagnets when creating interfaces with superconductors and can also bear important consequences for creation of altermagnetic Josephson junctions. Another promising direction for future study is to explore the interface between altermagnetism and $d$-wave superconductivity. It may also be interesting to study the behavior of altermagnetic interfaces with more exotic forms of superconductivity, such as the Bose-Einstein condensate regime \cite{lewandowskiAndreevReflectionSpectroscopy2023} or gapless superconductors with segmented Fermi surface \cite{zhuDiscoverySegmentedFermi2021}. With multitude of new research avenues, altermagnet/superconductor heterostructures promise fascinating phenomena, both from the basic research perspective as well as for potential technological applications.

\textit{Note added}.---
After the article submission we were made aware of a very recent related work on the Andreev reflection in altermagnets \cite{sunAndreevReflectionAltermagnets2023}, with the results in common between our works in full agreement.

\begin{acknowledgements}
 We thank Marc Vila Tusell for bringing our attention to the topic of altermagnets and Cyprian Lewandowski for helpful discussions. This material is based upon work supported by the U.S. Department of Energy, Office of Science, National Quantum Information Science Research Centers, Quantum Science Center.  M.P. received additional fellowship support from the Emergent Phenomena in Quantum Systems program of the Gordon and Betty Moore Foundation.
\end{acknowledgements}

\bibliography{altermagnetic_andreev}

\newpage
\onecolumngrid
\setcounter{equation}{0}
\setcounter{figure}{0}
\setcounter{page}{1}
\renewcommand{\thefigure}{S\arabic{figure}}
\renewcommand{\theequation}{S\arabic{equation}}

\begin{center}
\textbf{\large Supplemental Material for ``Andreev reflection at the altermagnet-superconductor interface"}

\,

{\large Micha{\l} Papaj}

\textit{Department of Physics, University of California, Berkeley, CA 94720, USA}
\end{center}

\section{Details for the analytical solution of the scattering problem at the altermagnet-superconductor interface}

As mentioned in the main text, care has to be taken when dealing with spatially varying coefficient terms in the Hamiltonian. This was recognized in particular in the field of modeling semiconductor heterostructures, where interfaces between materials described by different effective mass were considered. For example, when momentum operators are converted to differential operators, the standard kinetic term:
\begin{equation}
    T = \frac{\hbar^2(k_x^2 + k_y^2)}{2m^*(x)} = -\frac{\hbar^2}{2 m^*(x)} \nabla^2
\end{equation}
is not Hermitian when effective mass $m^*(x)$ takes an abrupt step at the boundary between the materials. Several modifications to restore hermiticity were proposed \cite{vonroosPositiondependentEffectiveMasses1983, morrowModelEffectivemassHamiltonians1984}, with the most common approach being converting the momentum operator along the $x$ direction to the differential operator:
\begin{equation}
    T_x = -\frac{\hbar^2}{2} \partial_x \frac{1}{m^*(x)} \partial_x
\end{equation}

Similar issues arise when converting the kinetic part in the current work that describes the anisotropic dispersion of the electrons with the opposite spin orientation. In such case, the differential operators used in the calculations are:

\begin{equation}
\label{eq:SM_kinetic}
    T = -\partial_x t(x) \partial_x - t(x)\partial_y^2 - (t_{J1}(x)\partial_y^2 - \partial_x t_{J1}(x) \partial_x) \sigma_z - (\partial_x t_{J2}(x)\partial_y + \partial_y t_{J2}(x) \partial_x)\sigma_z
\end{equation}
where $t(x) = t_0\theta(-x) + t_{SC}\theta(x)$, $t_{J1}(x) = t_{J1}\theta(-x)$, and $t_{J2}(x) = t_{J2}\theta(-x)$, with $\theta(x)$ being the Heaviside step function.

The abrupt change in the coefficients of the differential operators also warrants caution when discretizing it on the square lattice to perform numerical simulations. When transforming the continuum Hamiltonian to the lattice version with the lattice constant $a$, the general procedure was to approximate the derivative by the central difference:
\begin{equation}
    \partial_x f(x) \approx \frac{1}{a} \left(f(x + \frac{a}{2}) - f(x-\frac{a}{2}) \right)
\end{equation}
and recursively applying it to obtain higher order derivatives.

To establish the stitching condition for the wave function at the interface we can take the approach of integrating the Schrodinger equation with the BdG Hamiltonian within the infinitesimal distance $\epsilon$ from the boundary:
\begin{equation}
    \int_{-\epsilon}^{\epsilon} H_\mathrm{BdG} \Psi(\mathbf{r}) = \int_{-\epsilon}^{\epsilon} E\Psi(\mathbf{r})
\end{equation}
in which case the proper form of kinetic energy differential operator from Eq.~\eqref{eq:SM_kinetic} plays the crucial role and the result is given in the main text. We can now derive the quasiparticle current operator from its definition established through the continuity equation:
\begin{equation}
    \frac{\partial \rho}{\partial t} + \nabla \cdot \mathbf{j}(\mathbf{r}) = 0, \quad \quad \rho = \Psi^\dagger \Psi
\end{equation}
Calculation yields the $x$ component of the current operator of the following form in the altermagnet and the superconductor:
\begin{equation}
\label{eq:SM_current_op}
    j_x^{AM} = 2\, \mathrm{Im} \left[\Psi^\dagger(\mathbf{r}) \left(\tau_z(t_0 - t_{J1}\sigma_z)\partial_x + t_{J2}\partial_y)\right)\Psi(\mathbf{r})\right] , \quad j_x^{SC} = 2\, \mathrm{Im}\left[\Psi^\dagger(\mathbf{r})\left(t_{SC} \tau_z \partial_x \right)\Psi(\mathbf{r})  \right]
\end{equation}
The boundary condition presented in the main text thus guarantees continuity of the current along the $x$ direction. This in turn ensures that the scattering matrix that connects the wave function coefficients at the interface is unitary.

The Andreev and normal reflection amplitudes are then derived by solving the boundary matching condition separately for incoming spin up and spin down electrons. Such separation is possible because in the absence of spin-orbit coupling, spin is still a good quantum number in the altermagnet. The scattering states that we consider have the following form:
\begin{equation}
    \Psi_\uparrow(\mathbf{r}) = 
    \begin{cases}
    \left(e^{i k_{e\uparrow +} x + i k_y y} + r_{N\uparrow} e^{i k_{e\uparrow -} x + i k_y y}\right) (1, 0, 0, 0)^T + r_{A\uparrow} e^{i k_{h\downarrow -} x + i k_y y} (0, 0, 0, 1)^T & x < 0 \\
    a_\uparrow e^{i k_{SC e} + i k_y y} (E + \sqrt{E^2 - \Delta^2}, 0, 0, \Delta)^T + b_\uparrow e^{i k_{SC h} + i k_y y} (E - \sqrt{E^2 - \Delta^2}, 0, 0, \Delta)^T  & x > 0
    \end{cases}
\end{equation}
\begin{equation}
    \Psi_\downarrow(\mathbf{r}) = 
    \begin{cases}
    \left(e^{i k_{e\downarrow +} x + i k_y y} + r_{N\downarrow} e^{i k_{e\downarrow -} x + i k_y y}\right) (0, 1, 0, 0)^T + r_{A\downarrow} e^{i k_{h\uparrow -} x + i k_y y} (0, 0, 1, 0)^T & x < 0 \\
    a_\downarrow e^{i k_{SC e} + i k_y y} (0, -(E + \sqrt{E^2 - \Delta^2}), \Delta, 0)^T + b_\downarrow e^{i k_{SC h} + i k_y y} (0, -E + \sqrt{E^2 - \Delta^2}, \Delta, 0)^T  & x > 0
    \end{cases}
\end{equation}
The wave vectors are appropriately defined as:
\begin{equation}
    k_{\tau_z \sigma_z \pm} = \left(\pm \sqrt{(t_0 - \sigma_z t_{J1})(\mu + \tau_z E) - (t_0^2 - t_{J1}^2 - t_{J2}^2)k_y^2} - \sigma_z k_y t_{J2} \right) / (t_0 - \sigma_z t_{J1})
\end{equation}
\begin{equation}
    k_{SC\tau_z} = \tau_z \sqrt{\frac{\mu_{SC} + \tau_z \sqrt{E^2 - \Delta^2}}{t_{SC}} - k_y^2}
\end{equation}
where $\tau_z = e/h = \pm 1$, $\sigma_z = \uparrow/\downarrow = \pm 1$. Each of the components of the wave function is normalized to carry the same quasiparticle current using Eq.~\eqref{eq:SM_current_op} and then the equations arising from the boundary matching are solved for the reflection amplitudes. When there is no tunneling barrier at the interface and when we approximate $k_{SC\tau_z}\approx \tau_z \sqrt{\mu_{SC}/t_{SC}-k_y^2}$, we obtain the formulas for reflection amplitudes presented in the main text. However, all the plots were obtained using the full unapproximated amplitudes. If a delta tunneling barrier is present at the interface, the boundary condition includes additional $U_B \Psi(x=0)$ term that contributes to the derivative discontinuity. With such a term taken into account, the reflection coefficients are given by:
\begin{equation}
    r_{A\sigma} = \frac{2 \sigma \tilde{k}_{F,SC} \sqrt{\tilde{k}_{e\sigma}\tilde{k}_{h\bar{\sigma}}}}{\tilde{k}_{F,SC}(\tilde{k}_{e\sigma}+\tilde{k}_{h\bar{\sigma}})\tilde{\epsilon}+(\tilde{k}^{2}_{F,SC}+\tilde{k}_{e\sigma}\tilde{k}_{h\bar{\sigma}} - i\,U_B(\tilde{k}_{e\sigma}-\tilde{k}_{h\bar{\sigma}}) + U_B^2)\sqrt{\tilde{\epsilon}^2-1}} \\
\end{equation}
\begin{equation}
    r_{N\sigma} = \frac{\tilde{k}_{F,SC}(\tilde{k}_{e\sigma}-\tilde{k}_{h\bar{\sigma}})\tilde{\epsilon}+(\tilde{k}_{e\sigma}\tilde{k}_{h\bar{\sigma}}-\tilde{k}^{2}_{F,SC} - i\,U_B(\tilde{k}_{e\sigma}+\tilde{k}_{h\bar{\sigma}}) - U_B^2)\sqrt{\tilde{\epsilon}^2-1}}{\tilde{k}_{F,SC}(\tilde{k}_{e\sigma}+\tilde{k}_{h\bar{\sigma}})\tilde{\epsilon}+(\tilde{k}^{2}_{F,SC}+\tilde{k}_{e\sigma}\tilde{k}_{h\bar{\sigma}} - i\,U_B(\tilde{k}_{e\sigma}-\tilde{k}_{h\bar{\sigma}}) + U_B^2)\sqrt{\tilde{\epsilon}^2-1}}
\end{equation}
It is also possible to obtain analytical solutions for the case when the barrier is not at the interface, but placed somewhere inside the altermagnetic region. However, for the sake of comparison with the numerical calculation that contains disorder, in the main text we presented only the results obtained through the quantum transport simulation.

\section{Additional results for transport properties of the altermagnet-superconductor interface}
\begin{figure*}
    \centering
    \includegraphics[width=0.999\linewidth]{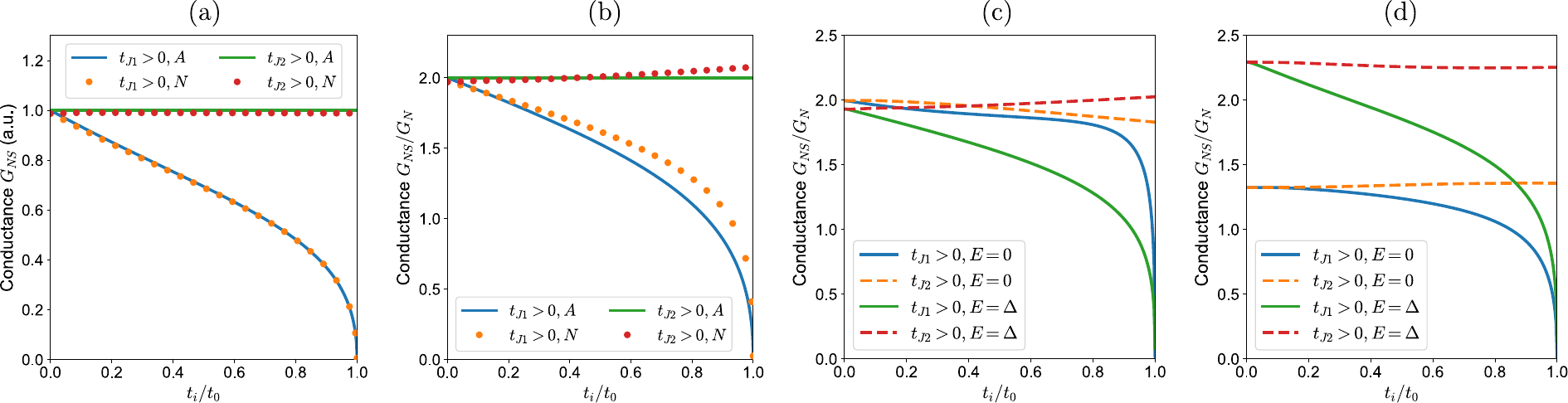}
    \caption{Dependence of altermagnet/superconductor interface conductance on the altermagnet parameters $t_i = t_{J1}, t_{J2}$. (a) Comparison between the analytical ($A$) and numerical ($N$) results at $E=\Delta$ for $G_{NS}$ as given by Eq.~\eqref{eq:cond_tJ1} for $t_{J1}>0$ and by $8\sqrt{\mu/t_0}$ for $t_{J2}>0$. (b) Comparison of the analytical and numerical results at $E=\Delta$ for the ratio $G_{NS}/G_N$. The discrepancy between the two approaches comes from the approximation for the normal state conductance of the interface in the analytical approach. (c) Numerical results at $E=0, \Delta$ without a barrier ($U_B=0$) at the interface. (d) Numerical results at $E=0, \Delta$ with a weak barrier ($U_B=0.625$) at the interface. In all cases, $t_0 =4$, $\mu=\mu_{SC}=0.25$, $\Delta = 0.001$.}
    \label{fig:SM_AM_parameters}
\end{figure*}

In addition to the results presented in the main text, we can also further characterize the dependence of conductance through the AM/SC interface as the parameters describing the altermagnet are varied. In the special case of $E=\Delta$, the integral for the conductance in the superconducting state can be evaluated analytically when either $t_{J1}$ or $t_{J2}$ are non-zero. The first case yields in the limit $\mu \gg \Delta$:
\begin{equation}
    G_{NS} \sim \frac{32}{15 t_{J1}^2} \sqrt{\frac{\mu}{t_+t_-}} \left(t_-^{\frac{3}{2}}(t_- - 5t_+) + \sqrt{t_+} \left( (4t_0^2 - 3t_{J1}^2) E\left(\frac{t_-}{t_+}\right) + t_{J1}(3t_{J1} -4t_0) K\left(\frac{t_-}{t_+}\right) \right) \right)
    \label{eq:cond_tJ1}
\end{equation}
where $t_\pm = t_0 \pm t_{J1}$, and $K(x), E(x)$ are complete elliptic integrals of the first and second kind, respectively. The second case where $t_0>t_{J2}>0$ gives approximately a constant result $G_{NS}\sim8\sqrt{\mu/t_0}$. The comparison between both formulas and the numerical integration is presented in Fig.~\ref{fig:SM_AM_parameters}(a), which shows a very good agreement. To relate it to the conductance in normal state, we can approximate $G_N$ for spin up and down components in the $t_{J1}$ case as $2\sqrt{\mu/t_+}$ and $2\sqrt{\mu_{SC}/t_{SC}}$, respectively, assuming $t_0>t_{J1}>0$. In the $t_{J2}$ case, both spin components have equal contributions equal to $2\sqrt{\mu_{SC}/t_{SC}}$. With such estimates, we can calculate $G_{NS}/G_N$ ratio, which is the quantity presented in the main text. The result is presented in Fig.~\ref{fig:SM_AM_parameters}(b). While the approximation for $G_N$ slightly overestimates its value, leading to the underestimation of the $G_{NS}/G_N$ ratio as compared to the true result computed numerically, this approach reproduces all the qualitative features of this dependence. For $t_{J1}$ case, this means a steady decrease for small $t_{J1}$, with a fast drop to 0 close to $t_0=t_{J1}$ point, while in $t_{J2}$ case the conductance ratio remains roughly constant throughout the parameter range.

To complete this analysis, we present the numerical results for both $E=0$ and $E=\Delta$ cases without (Fig.~\ref{fig:SM_AM_parameters}(c)) and with (Fig.~\ref{fig:SM_AM_parameters}(d)) tunnelling barrier placed at the interface. In both cases, the difference between the two orientations of the Fermi surface is significant: while for $t_{J1}>0$ the conductance systematically decreases as $t_{J1}$ increases, with a very rapid drop close to $t_{J1}=t_0$, for $t_{J2}>0$ the conductance remains roughly constant even though the spin splitting increases significantly. This is because for $t_{J2}$ Fermi surface orientation the number of spin up electron and spin down hole states (and vice versa) remains the same so that the Andreev reflection for both components is not suppressed. The conductance decrease rate in the $t_{J1}$ case increases for $E=\Delta$ compared to $E=0$, but the qualitative features remain the same. This characterization is also independent of the presence of the barrier, which mostly determines the initial value of $G_{NS}/G_N$ ratio.

Finally, we can consider how the parameters used in the simple model can be related to the real material parameters. For example, in metallic altermagnet RuO$_2$ the band widths of the spin-split bands are on the order of 1 eV, while the spin splitting itself can be larger than 500 meV \cite{smejkalEmergingResearchLandscape2022}. When translated to the $t_0$ and $t_{J}$ parameters this enables us to establish the ratio to be about $t_J/t_0 \approx 0.5$. However, materials with larger relative spin-splitting exist, which could allow $t_J$ values even closer to $t_0$. Nevertheless, a proper investigation of transport properties of a junction with a specific material would require a more detailed model of a given system, preferably with a band structure based on the first principles calculations.

\end{document}